# Pressure-induced superconductivity in topological type II Dirac semimetal NiTe₂

Tao Li[1], Ke Wang[1,2], Chunqiang Xu[1],Qiang Hou[1], Hao Wu[1,2], Jun-Yi Ge[1], Shixun Cao[1], Jincang Zhang[1], Wei Ren[1], Xiaofeng Xu[3], Nai-Chang Yeh[4], Bin Chen[2], Zhenjie Feng[1, 5*]

*[1]Materials Genome Institute, Department of Physics, Shanghai University, Shanghai, 200444, China*

*[2]Center for High Pressure Science and Technology Advanced Research, Shanghai 201203, China*

*[3]Department of Physics, Changshu Institute of Technology, Changshu 215500, China*

*[4]Department of Physics, California Institute of Technology, Pasadena, CA 91125, USA*

*[5]Shanghai Key Laboratory of High Temperature Superconductors, Shanghai 200444, China*

## Abstract

Very recently, NiTe₂ has been reported to be a type II Dirac semimetal with Dirac nodes near the Fermi surface. Furthermore, it is unveiled that NiTe₂ presents the Hall Effect, which is ascribed to orbital magnetoresistance. The physical properties behavior of NiTe₂ under high pressure attracts us. In this paper, we investigate the electrical properties of polycrystalline NiTe₂ by application of pressure ranging from 3.4GPa to 54.45Gpa. Superconductivity emerges at critical pressure 12GPa with a transition temperature of 3.7K, and Tc reaches its maximum, 7.5 K, at the pressure of 52.8GPa. Comparing with the superconductivity in MoP, we purposed the possibility of topological superconductivity in NiTe₂.

## Introduction

In recent years, The transition metal dichalcogenides (TMDs) has been extensively investigated for their plentiful physic properties. As a special part of TMDs, MX₂ (M=Mo, W, Zr, Pt, Pd, Ni; X=Te) attract growing attention in condensed-matter physics. Most of them can achieve superconductivity by doping or pressure. The intrinsic superconductivity under 0.1K[1] in Weyl semimetal candidate MoTe₂ can be enhanced by partial replacing the Te with S[2] and Se[3](Tc can increase to 1.2K and 2.5K for S and Se substitution respectively). Besides, the application of pressure also facilitates the superconductivity in MoTe₂[1], which establishes a dome-like pressure phase diagram. For WTe₂, superconductivity emerges by applying exotic pressure accompanied by disappearance of large magneto-resistance state[4]. In contrast to pressure, intercalation of potassium induces superconductivity in WTe₂ without suppressing the MR[5]. Furthermore, Cu$_x$ZrTe$_{2-y}$ is reported to be superconductors containing topological properties[6], and PtTe₂, PdTe₂ [7, 8]are wildly studied as type II Dirac/Weyl semimetal candidates.

In comparison, NiTe₂ attract less attention until latest two years. Before that, major researches of NiTe₂ are about crystal structure or band structure[9, 10]. Until 2018, linear field dependence of magnetoresistance was discovered in NiTe₂ which originates from Dirac nodes near the fermi surface[11], making the Dirac quasi-particles dominate the physical behavior in NiTe₂. In addition, NiTe₂ has been reported to exhibit planar Hall effect attributed to orbital magnetoresistance[12]. Rich physics properties make NiTe₂ a potential material deserved more investigation.

It is well accepted that pressure is an effective method to modify electronic structure of semiconducting materials thereby inducing superconductivity, especially for materials which undergo magnetic transition under ambient pressure or possess plentiful physic properties. For instance, pressure can facilitate the growing of superconducting state by suppressing the anti-ferromagnetic state in CrAs. In this case, superconductivity occurs at critical pressure where MR has been vanished. To this respect, it is of great possibility to achieve superconductivity in NiTe₂ resorting to pressure method.

In this paper, we investigate the pressure impression on resistant behavior of NiTe₂. NiTe₂ presents metal property at ambient pressure in temperature ranging from 2K-300K. Superconductivity advents at 12GPa, with transition temperature of 3.7K. The Tc is enhanced by applying exotic pressure, which reaches to 7.5K at 52.8GPa. Further increasing pressure leads a slight dwindling of Tc. Comparing with the pressure-induced superconductivity in MoP[13], we purpose the possibility of topological superconductivity in NiTe₂.

**Sample preparation**

Polycrystalline sample of NiTe₂ was synthesized by solid-state reaction method. Pure elements Ni (99.99%), Te (99.99%) were weighted with molar ratio of 1:2 in Ar₂-filled glove box. Then the starting material was grounded and pelletized into pellets before sealed into an evacuated silica tube. The tube containing precursor was placed in furnace heated at 750℃ for 48h followed by a second heat treatment in same condition after reground and pelletizing process. Eventually, the sample was quenched in cold water in order to gain a better crystallization of NiTe₂. The obtained sample presents metallic gloss with monocrystal-like small size flakes forming inside.

**Experimental details**

Power x-ray diffraction measurement under ambient pressure was carried out on Bruker D2 phaser with Cu κₐ radiation (λ=1.5418Å). The chemical composition was measured on scanning electron microscope SU5000 equipped with EDX. Diamond anvil cell (DAC) was employed to generate pressure on the power sample and pressure were determined by ruby power using ruby fluorescence method. Resistant measurement was performed on physics properties measurement system (PPMS 14) resorting four-probe method.

**Result and discussion**

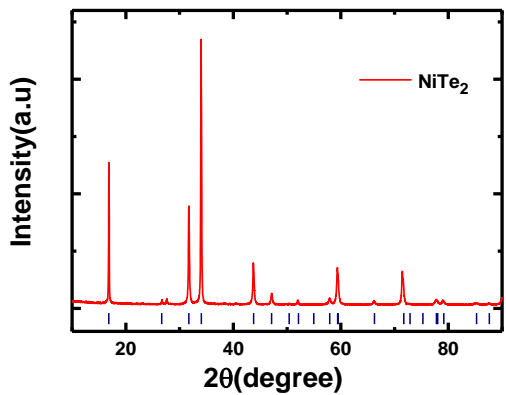

Figure 1. The XRD pattern of polycrystalline NiTe₂.

Powder XRD pattern of NiTe₂ is shown in Figure 1(a).

Almost all the pattern peaks can be indexed by trigonal structure with *P-3m1* space group. The obtained lattice parameters a, b, c are 3.85986 Å, 3.85986 Å, 5.28187 Å, respectively, which are similar to ones reported in Ref[11]. We resort to Energy-dispersive x-ray spectroscopy to confirm the certain composition in NiTe₂ and the result was shown in Figure 1(b). The average molar ratio of Ni and Te element collected from 6 points of power sample is 38:62 roughly close to the stoichiometric composition 33:66.

Figure 2(a) shows the temperature dependence of resistance under various pressure at zero magnetic field. All the resistance data present almost linear diminishing with decreasing temperature at high temperature region regardless different applied pressure, taken as a characteristic of metallic behavior. Under lower pressure (≤18.1GPa), the resistance curve almost overlaps and an anomaly appears under 12.0GPa shown as a slight drop of resistance at low temperature nearly 3K, which is confirmed to be the signature of superconductivity by resistance data under magnetic field. Upon further increasing the pressure, the whole R-T curve at normal state is upshifted by a constant remaining the same temperature dependence picture, which is reminiscent of pressure studies on Weyl semimetal candidate MoP[13]. MoP is experimentally proved to contains Weyl points near the fermi surface analogous to Dirac points in NiTe₂ and the pressure-induced superconductivity in MoP offers a promising candidate of topological super-conductivity supported by evidence from structure measurement under pressure and ab initio theoretical calculation.

Back to the case of NiTe₂, at higher pressure region the phenomenon of resistance dropping in low temperature becomes much more prominent with increasing pressure. The amplitude of drop enhances and Tc$^{onset}$ defined as the temperature point where resistance starts to drop simultaneously increases. At pressure of 52.8GPa, Tc$^{onset}$ grows to its maximum of 7.5K. We did second run of electric resistance measurement at two selected pressure point of 3.4GPa and 54.45GPa and the data is shown in Figure 2(b). The resistance presents no anomaly at low temperature region under 3.4GPa while in 54.45GPa the dropping feature is enhanced with only slightly shrinking of Tc comparing to 52.8GPa. Within this tendency, it is of considerable possibility to realize zero-electrical resistance in NiTe₂ upon higher pressure. It is worth noted that no anomaly has been discovered in normal state region under 3.4GPa and 54.45 GPa in agreement with date from run 1. Figure 2(c) shows the resistance date under different pressure at certain temperature extracted from date in Figure 2(a). The pressure dependence of resistance exhibits similar behavior at selected temperature points. The resistance is insensitive to pressure at low pressure region then raises monotonically by increasing pressure. At lower temperature, the resistance value at 18.07GPa shifts upward continuously thereby making the pressure independent character of resistance at low pressure less notable eventually resulting in a much more smooth R-P curve of 50K. The data of lower temperature will render the insight of actual properties in material. From the R-P curve at 50K, the anomaly at high temperature is almost ruled out indicating absent of structure transition. Interestingly, the MoP exhibits similar pressure dependence of resistance in the normal state. In contrast, for many other topologic materials, the pressure-induced superconductivity is often concomitant with structure transition manifested as an anomaly in normal state R-P curve, which will break the crystal symmetry leading to the disappearance of the topologic feature[14-16]. The absence of anomaly in pressure dependence of normal state resistance simply denies the existence of structure transition, ensuring the possibility for topological nature of pressure-induced superconductivity in NiTe₂. In many previous works of pressure-induced superconductivity, the temperature dependence of normal state resistance near the superconducting transition often follows the relation formulated as $\rho = \rho_0 + AT^n$. The A is characteristic for different materials and n is exponent, playing an significant role to understanding the mechanism in pressure-induced superconductivity. In the pressure studies of CrAs[17] an TiSe$_2$[18], the exponent n is reported to notably decreased in the pressure region where superconductivity occurs. In our case, the $\rho$ is replaced by R and the equation changes to be $R = R_0 + AT^n$. We derive the exponent n by fitting the resistance date from Figure 2(a) in temperature region from 10K to 70K and the result is shown in Figure 2(d). The n hovers around the value~2.0 under distinct pressure, regarded as a signature of Fermi liquid metals. The pressure-independent n indicates that electron-electron interaction is almost unchanged upon pressure.

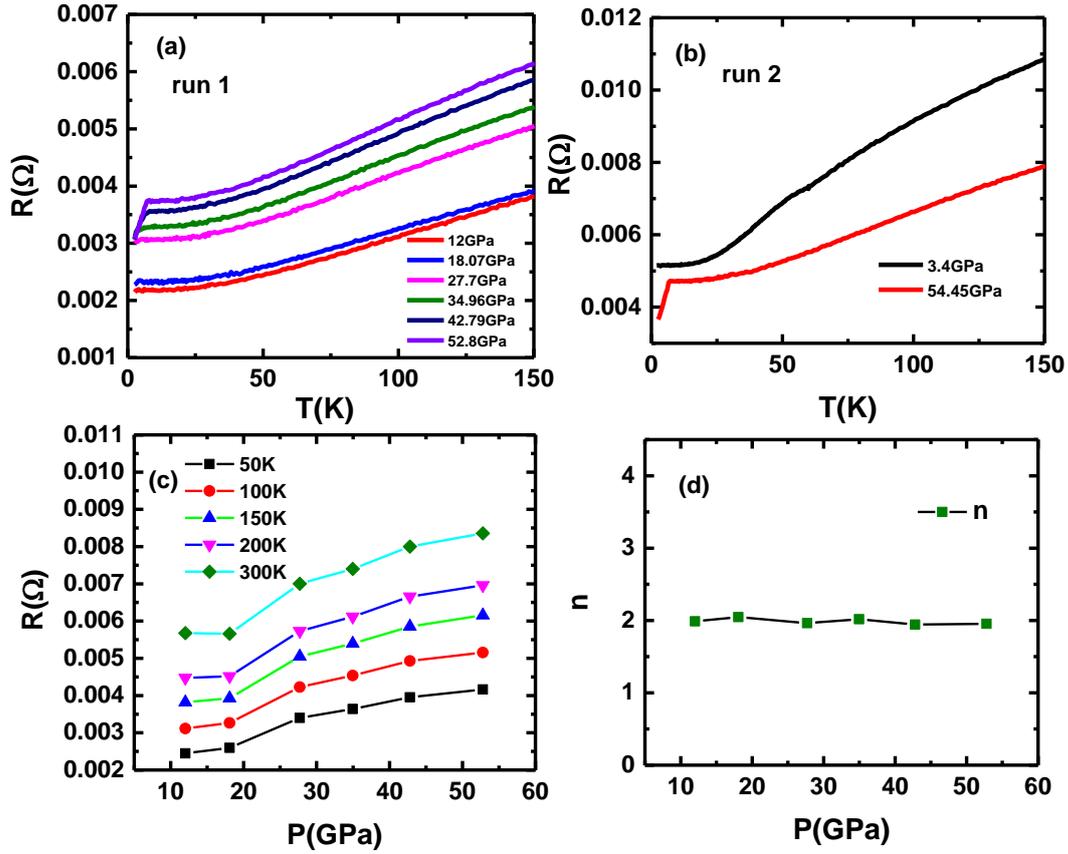

Figure 2. (a) Temperature dependence of resistance from 2K to 150K under various pressures for run 1. (b) The resistance data at 3.4 GPa and 54.45 GPa of run 2. (c) The Resistance vs pressure at fixed temperature. (d) The exponent n obtained by fitting the resistant data to $R = R_0 + AT^n$ from 10K to 70K.

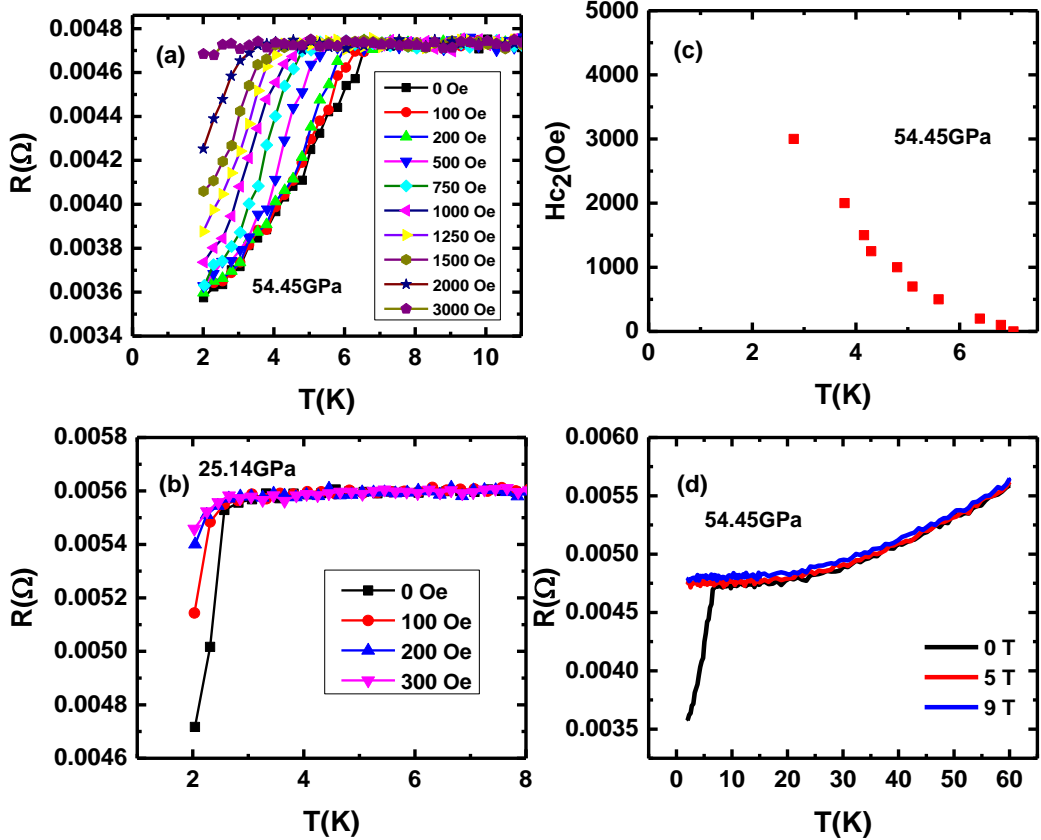

Figure 3 (a) Temperature dependence of resistance under distinct magnetic field for 54.45GPa at low temperature region. (b) The resistance data under field for 25.14GPa. (c) The Hc2 vs T plot at 54.45GPa. (d) The magnetoresistance behavior under pressure of 54.45GPa.

We further measured the electrical resistance of NiTe$_2$ under magnetic field to confirm the existence of superconductivity. Figure 3(a) shows the resistance behavior at 54.45GPa under various field. With enlarging the applied field, the phenomenon of resistant dropping is gradually suppressed and consequently vanishes at 3000 Oe accompanied by decreasing of Tc$^{onset}$, verifying the anomaly is originated from superconductivity transition. We repeated the same procedure at pressure of 25.14GPa, deriving the identical result, which further confirm the truth. In Figure (c), the upper critical field (Hc$_2$) of NiTe$_2$ exhibits a positive curvature deviating from the Werthamer-Helfand-Hohenberg theory based on the single-band model. This deviation is reminiscent of the case for MgB$_2$, NbSe$_2$[19] and Cu$_x$ZrTe$_{2-y}$[6], taken as a signature of multiband behavior. It should be mentioned that Hall resistivity studies at ambient pressure demonstrates the multiband electrical transport channels in NiTe$_2$, which is consistent with theoretical calculation. The pressure-induced superconductivity remains multiband feature even under such high pressure of 54.45GPa. As shown in Figure 3(c), our experiment curve of Hc$_2$(T) can be well fitted by relation equation $H_{c2}(T) = H_{c2}(1 - T/T_c)^{1+\alpha}$ .

NiTe$_2$ was reported to exhibit linear magneto-resistance, which can reach as high as 1250% at 2K and 9T. In order to investigate the pressure influence on MR, we preformed high field resistance measurement for 54.45GPa. As shown in Figure 3(d), the resistance almost persists unchanged under 0T, 5T, and 9T, confirming the absence of magnetoresistance. We also carried out same measurement at 3.4GPa rendering the similar result, which was not shown here. In this consideration, the MR will be completely suppressed by small pressure as high as 3.4GPa. But it is of possibility that the single crystal sample of NiTe$_2$ behaves differently under pressure. Therefore, further pressure investigation on single crystal is required to obtain clear detail about the impact of pressure on MR. Finally, we gain the superconducting temperature-pressure phase diagram as shown in Figure 4. The superconducting transition occurs at 12.0 GPa with Tc$^{onset}$ ∼ 3.5 K. The Tc$^{onset}$ is enhanced monotonously by increasing the applied pressure and reaches 7.5 K at 52.8GPa eventually slightly decreases to 7.1 K at 54.45 GPa. It should be note that the phase diagram is very similar to that of MoP.

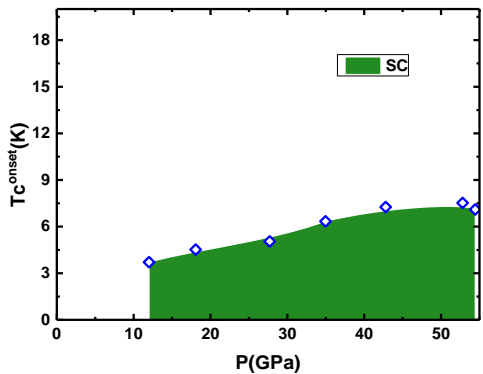

Figure 4. The pressure-superconductivity phase diagram of NiTe$_2$.

**Conclusion**

We investigate the pressure effect on electrical properties of polycrystalline NiTe$_2$ and realize superconductivity starting from 12GPa with Tc of 3.7K. Tc grows with enlarging the pressure and eventually reached its maximum value of 7.5K at 52.8GPa. We observe no anomaly in the pressure dependence of normal state resistance(at 50K), indicating the absence of structure transition under pressure. This ensures the possibility of topological nature of superconductivity in NiTe$_2$. Besides, the pressure-induced superconductivity maintains multiband feature similar to pure NiTe$_2$ under ambient pressure suggested by the Hc$_2$-T data.

High field studies demonstrate that MR state of polycrystalline sample can be smeared out by application of pressure about 3GPa, which maybe differs in single crystalline one. Strikingly, there are so much common feature between the NiTe$_2$ and MoP. They also contain the Dirac or Weyl nodes near the fermi surface, can present superconductivity under pressure and follow the similar resistance behavior upon pressure. The pressure-induced SC in NiTe$_2$ and MoP also exhibit multiband feature and Tc-pressure phase diagram renders in the similar shape. Finally, we claim that we discovered the pressure-induced superconductivity in NiTe$_2$. This work enriches the Ni-based superconductor family and this similarity with pressure-induced superconductivity in MoP providing more experimental evidence for further researches on topologic quantum materials which the Dirac or Weyl nodes is closed to the fermi surface.

**Acknowledgement:**
This work at the Shanghai University (SHU) is jointly supported by the Ministry of Science and Technology of the People's Republic of China No. 2016YFB0700201, National Natural Science Foundation of China (11774217, 10904088), Shanghai Pujiang Program (13PJD015), and Science and Technology commission of Shanghai Municipality (13ZR1415200). N.-C. Yeh acknowledges the hospitality and sponsorship of her visit to the SHU under the Overseas Expert Recruitment Program at SHU.
* fengzhenjie@shu.edu.cn

**Statement:** The manuscript is currently quite rough. The first version of the manuscript about the observation of superconductivity in powder sample under high pressure was finished in 29th June, 2019. Bin Chen provided the high pressure technology. Nai-chang Yeh provided meaningful discussion and suggest to redo experiments on single crystal samples. After that, we tried to redo the experiments and confirm its topological superconductivity in single crystal NiTe₂ samples. We have obtained the single crystal samples from Xiaofeng Xu and the measurement is in progress. On 12th November, Feipeng Zheng *et al*, released their calculation results of NiTe2 on arXiv (1911.04668v1). Their results show that "monolayer NiTe2 to be an intrinsic superconductor with a Tc ∼5.7 K, although the bulk crystal is not known to superconduct". Therefore, we decide to release our research results on arXiv immediately. The manuscript will be updated with new data from NiTe₂ single crystals, and the *ab initio* calculations by Prof Wei Ren ASAP. The authorship list will also be updated according to the new contribution to the manuscript.

**2019.12.22.New Results:**

We have finished the resistance and Hall measurements. The results are shown in the following figures. We found two superconductivity transition under pressure. The detailed analysis will be published soon.

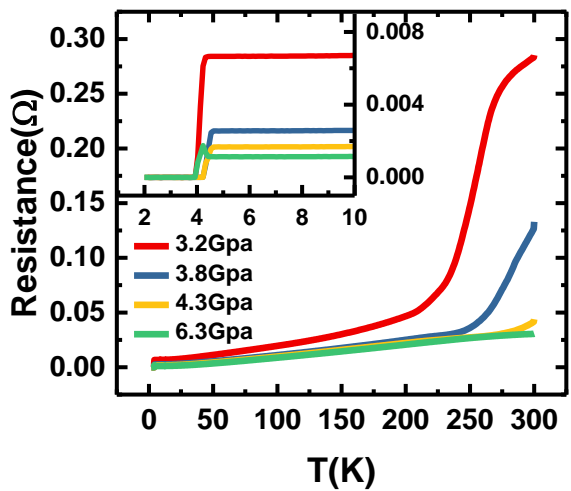

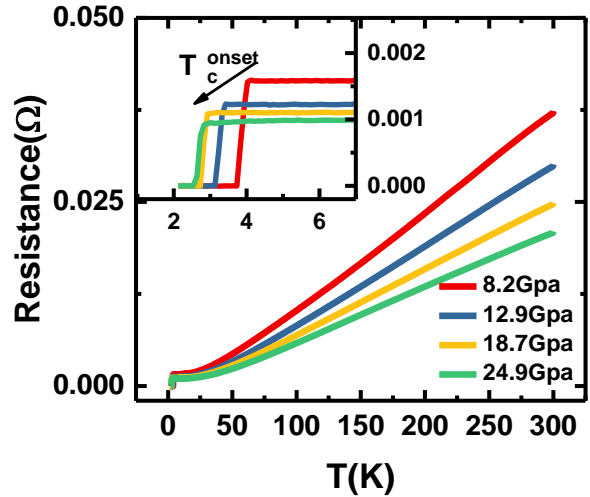

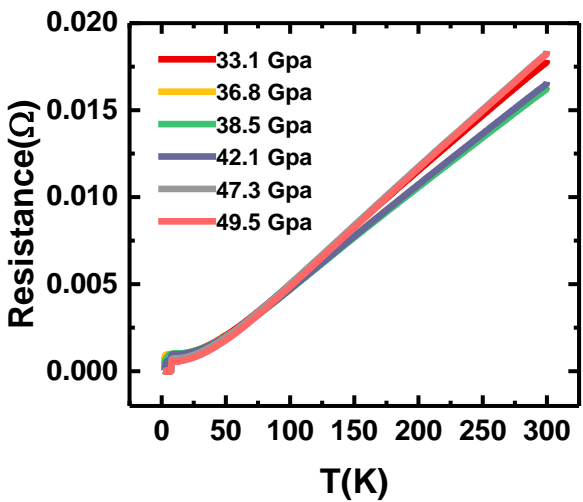

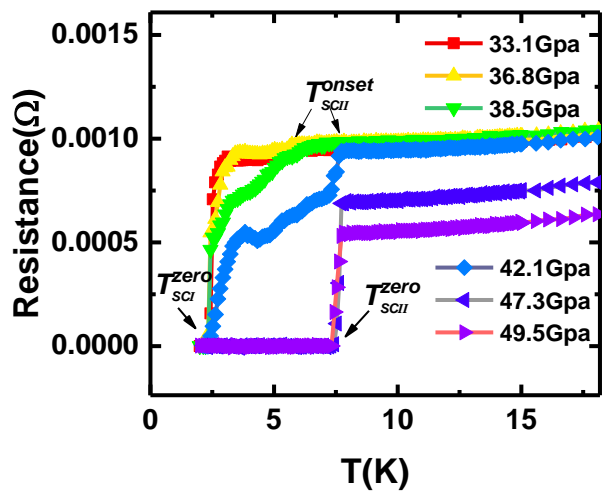

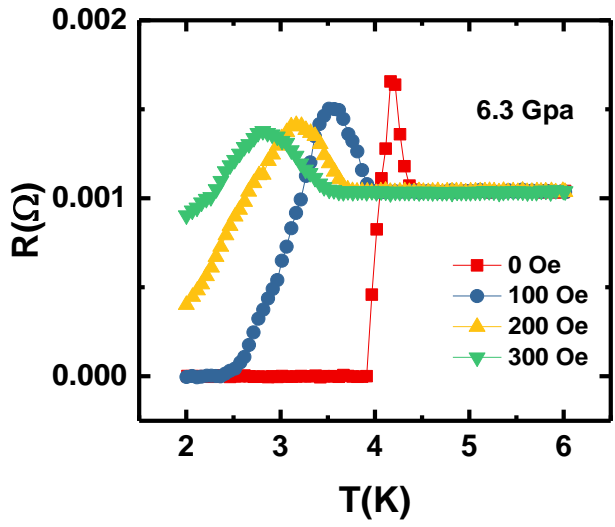

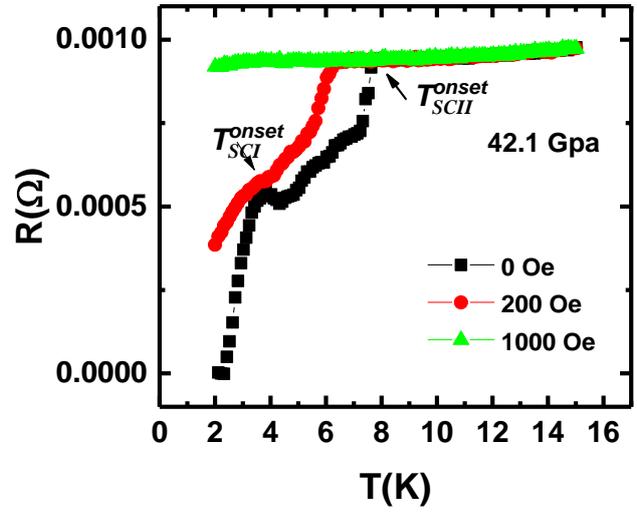

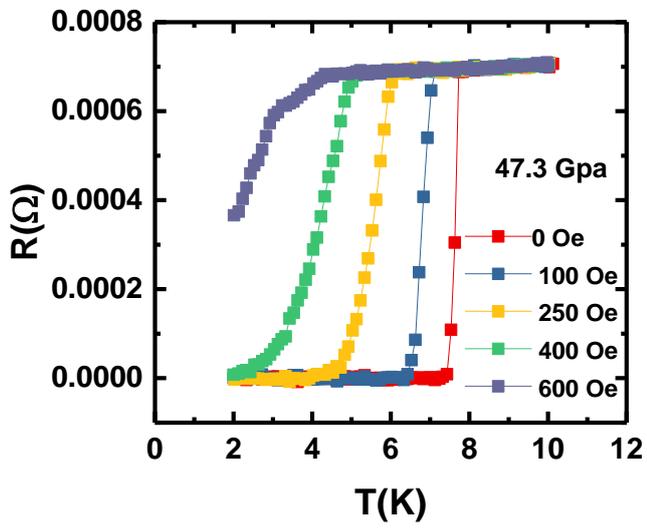

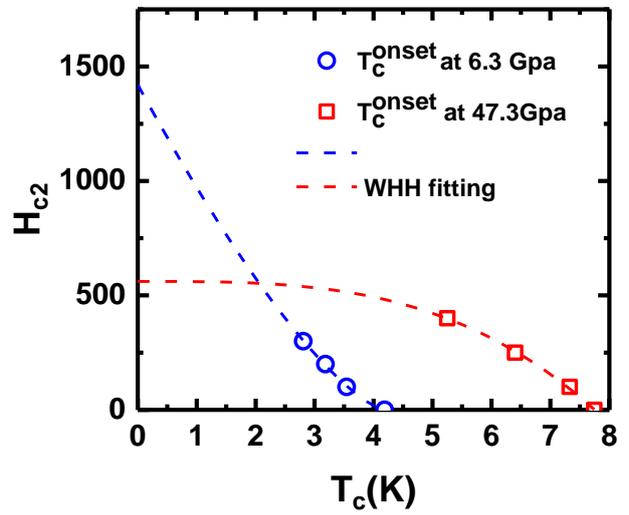

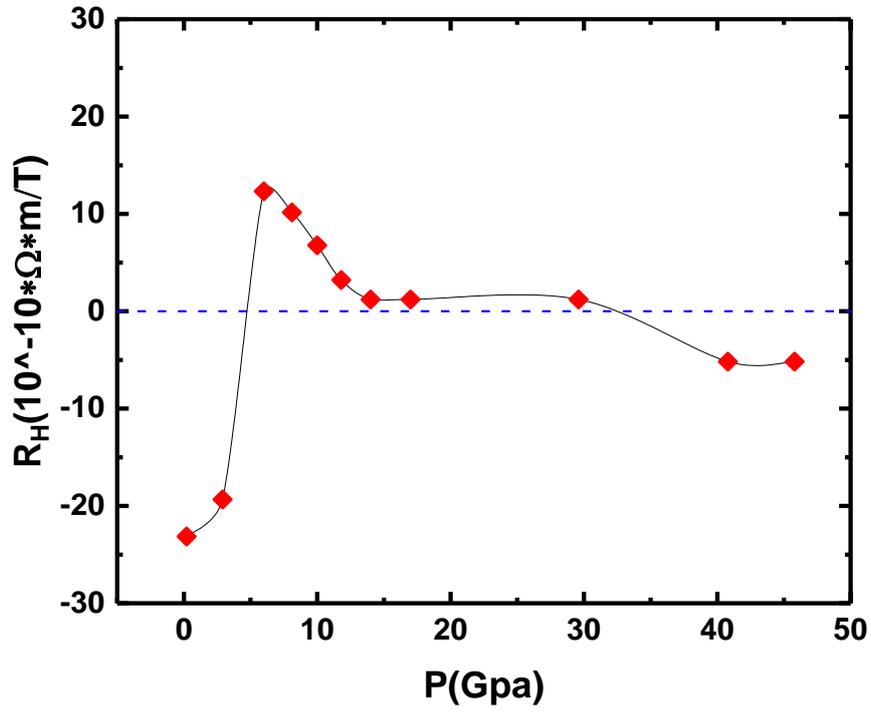

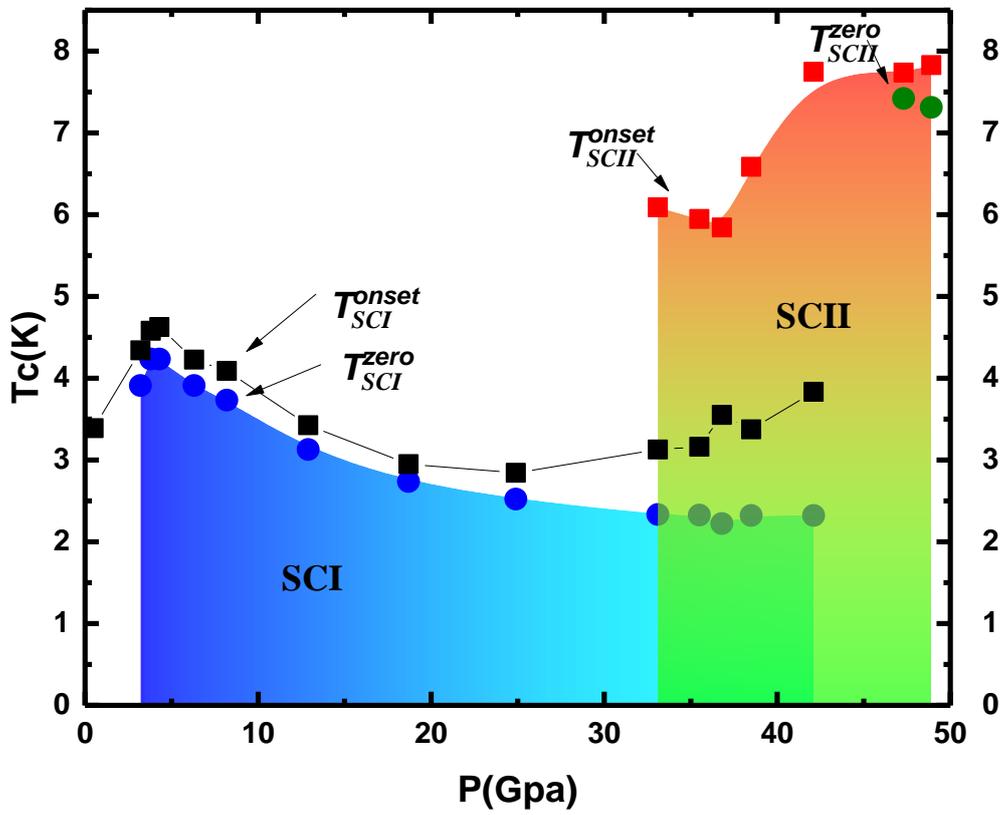